\documentclass[reprint,twocolumn,amssymb,nobibnotes,aps,prl]{revtex4-1}
\usepackage{graphicx}
\usepackage{epstopdf}

\begin{document}

\title{Light with tunable non-Markovian phase imprint}%

\author{Robert Fischer$^1$}
\email[Corresponding author: ]{r@robertfischer.eu}
\author{Itamar Vidal$^{2,4}$}
\author{Doron Gilboa$^4$}
\author{Ricardo R. B. Correia$^3$}
\author{Ana C. Ribeiro-Teixeira$^3$}
\author{Sandra D. Prado$^3$}
\author{Jandir Hickman$^3$}
\author{and Yaron Silberberg$^4$}
\affiliation{$^1$CPGEI, Federal University of Technology - Paran\'a, 80230-901 Curitiba, PR, Brazil \\$^2$Grupo de F\'isica At\^omica e Lasers, DF-CCEN, Universidade Federal da Para\'iba, Cx. Postal 5086, 58051-900 Jo\~ao Pessoa, Paraiba, Brazil \\$^3$ Instituto de F\'isica, Universidade Federal do Rio Grande do Sul, 91501-970 Porto Alegre, RS, Brazil  \\$^4$Department of Physics of Complex Systems, Weizmann Institute of Science, Rehovot 7610001, Israel}


\begin{abstract}

We introduce a simple and flexible method to generate spatially non-Markovian light with tunable coherence properties in one and two dimensions. The unusual behavior of this light is demonstrated experimentally by probing the far field and recording its diffraction pattern after a double slit: In both cases we observe instead of a central intensity maximum a line or cross shaped dark region, whose width and profile depend on the non-Markovian coherence properties. Since these properties can be controlled and easily reproduced in experiment, the presented approach lends itself to serve as a testbed to gain a deeper understanding of non-Markovian processes.
\end{abstract}

\maketitle

Although in nature, non-Markovian processes are far more common than Markovian ones, scientists often prefer to describe its behavior with Markovian models. This might be the case because Markovian models are easier to build, and significantly easier to combine when constructing more complex models. Recently, however, there is an increasing interest in understanding the specific impact of non-Markovian behavior on chemical and biological processes such as electron transfer~\cite{Xu_electronTransfer_2007}, kinetics of protein folding~\cite{Plotkin_ProteinFolding_1998}, or the ion transport through membranes~\cite{Goychuk_ionChannel_2003}, to name but a few. In optics, the radiation dynamics in photonic crystals~\cite{Hoeppe_nMRadiationInPC_2012}, optical gain in quantum-well lasers~\cite{Ahn_theoryQuantumWellLaser_1997}, and dephasing processes in coupled quantum-dot-cavities~\cite{Kaer_dephasingCQDCavity_2010} have been experimentally observed to be governed by non-Markovian behavior. Characterizing non-Markovian evolutions of quantum properties at the loss of entanglement has become the center of extensive theoretical and experimental efforts~\cite{Liu_expControlNonMarkov_2011, Chiuri_linOpticQuantumSim_2012}, since it is hoped to pave the way for better read-out mechanism in quantum computing~\cite{Goldstein_quantumPrecisionMeasurement_2011} or even controlled generation of entangled states~\cite{Cho_opticalPumpingEntanglement_2011}. Such quantum phenomena, or even their application in logical circuits~\cite{crespi_quantumOptCircuit_2011}, are often studied using optical means~\cite{Liu_expControlNonMarkov_2011,Chiuri_linOpticQuantumSim_2012}. Furthermore, the recent suggestion to initialize controlled quantum states by 'optical pumping'~\cite{Cho_opticalPumpingEntanglement_2011}, opens exciting possibilities to shape desired quantum properties, including designed non-Markovian characteristics, in the optical regime and then transfer it to other physical systems~\cite{Barreiro_pumpingTrappedIonsQuantumSim_2011,Paternostro_opticalCreatingNProbingEntanglement_2007}.

Non-Markovian light, which might serve as an effective tool towards these objectives, has been previously generated by overlaying a light beam with a delayed copy or echo of itself~\cite{Hartmann_echo_1984}. An alternative approach employs a micro-mechanical oscillator as a mirror to obtain non-Markovian properties from its Brownian movements~\cite{Groeblacher_BrownMirror_2013}. The two methods result in light with non-Markovian properties in the time domain and offer only a limited degree of control and practically no repeatability if the exact same conditions need to be reproduced. In this work we therefore introduce and experimentally demonstrate a technique to create \emph{spatially} non-Markovian light to overcome these limitations. By using a spatial light modulator (SLM) to imprint a non-Markovian phase pattern (NMP) onto a coherent light beam, we build a simple yet flexible, programmable and thus easily reproducible experimental setup to generate light with the desired spatial non-Markovian properties in one or two dimensions. Due to this combination of control over the non-Markovian properties and rapid experimental realization, this approach lends itself to serve as a testbed to gain a deeper understanding of the dynamics of non-Markovian processes, or even as a building bloc for experiments aiming at producing specific entangled states.

 \begin{figure}
 	\centering
 	\includegraphics[width=0.7\linewidth,clip=true]{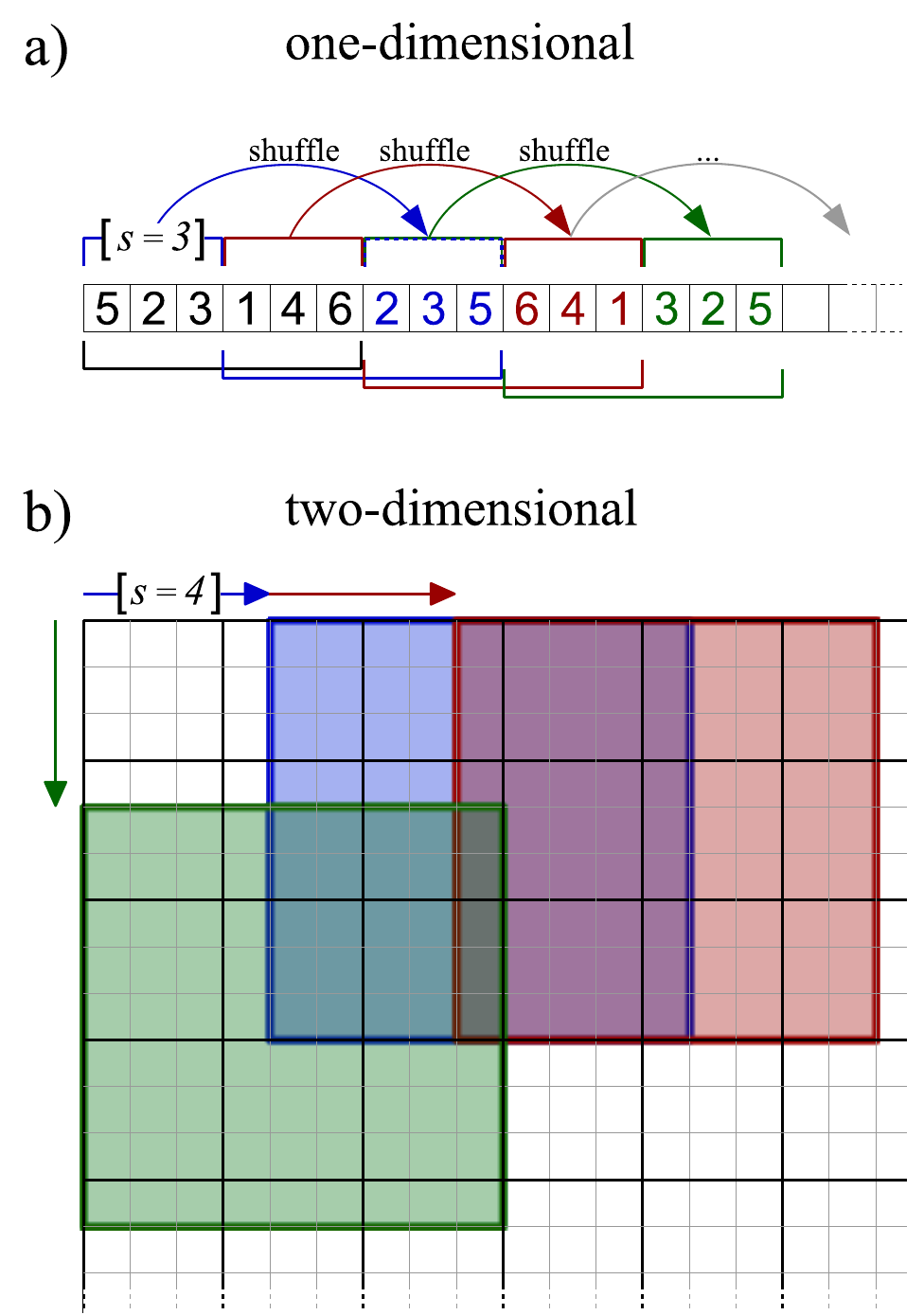}
 	\caption[Pattern generation]{Generation of non-Markovian patterns for phase imprint. a) In one dimension: generating a series of overlapping permutations (indicated by colored braces below series) of here $L=6$ symbols, e.g. with a shift size $s=3$. b) In two dimensions: Solving overlapping Sudoku puzzels after shifting the outer $9\times9$ frame by e.g. $s=4$ elements horizontally or vertically.}
 	\label{fig:patGen}
 \end{figure}

Moreover, NMPs in both one and two dimensions can be generated by algorithms with a tuning parameter that allows to scan from strictly periodic to non-Markovian properties. Looking at the spatial coherence of the wavefront shaped by this means, we observe, depending on the tuning parameter, a short-range anti-correlation, mid-range correlation, yet practically no correlation beyond a specific distance. Probing this light in two basic experiments, namely by recording the far field and diffraction on a double split, we demonstrate how the response of this light differs significantly form the familiar behavior of coherent or thermal light, or any combination of these two.

In a Markovian process, the probability of a given state may depend only upon the one state preceding it (relaxed condition), but not on the sequence of events that occurred before. Applying the Markov property to the case of a two-dimensional matrix, the value at each matrix element should be statistically independent from any other element with the exception of its immediate neighbors at either side. As a consequence, the covariance matrix of a pattern fulfilling the Markov criteria will be (tri)diagonal~\cite{horstmeyer_markov_2012}. However, the Markov condition is violated if the statistical process generating the pattern has some kind of memory. A textbook example would be the drawing of distinct balls from a urn without replacing, where the probability of a ball to be drawn depends on the outcome of all previous drawings. As shown in Fig.~\ref{fig:patGen}(a), we use a slightly modified version of this model to generate one-dimensional NMP: Starting with a random permutation of $L$ symbols (e.g. for $L=9$ the numbers 1~to~9), the series is continued with a random permutation of its first $s$ elements ($1\leq s \leq L$), such that the tuple of the last $L$ elements again form a permutation of $L$. Then the next $s$ elements in the series are added in randomly permutated order, and the process repeated until the series reaches the desired length. For our 1D experiments, we combine several such lines (each statistically independent from the others) to a matrix with horizontally non-Markovian, yet vertically Markovian properties.

One convenient method to generate two-dimensional NMPs is the solving of overlapping Latin squares or, alternatively, Sudoku puzzles. For this work we use common $9\times9$ Sudokus with the numbers 1 to 9 as symbols. Starting with one correctly solved Sudoku we shift it's outer frame of size $9\times9$ by $s$ elements either horizontally or vertically and fill the empty elements to solve for a valid Sudoku within the new frame position(Fig.~\ref{fig:patGen}(b)). The frame then is scanned in steps of $s$ elements over the matrix until the desired dimensions are reached. The number of $L=9$ symbols was chosen primarily for clarity in the reported figures and the familiarity with regular Sudokus; yet both in one and two dimensionen higher values can be used, thus increasing the range of $s$ and equally the resolution of the transition between strictly period and non-markovian patterns.

\begin{figure}
	\centering
	\includegraphics[width=0.9\linewidth,clip=true]{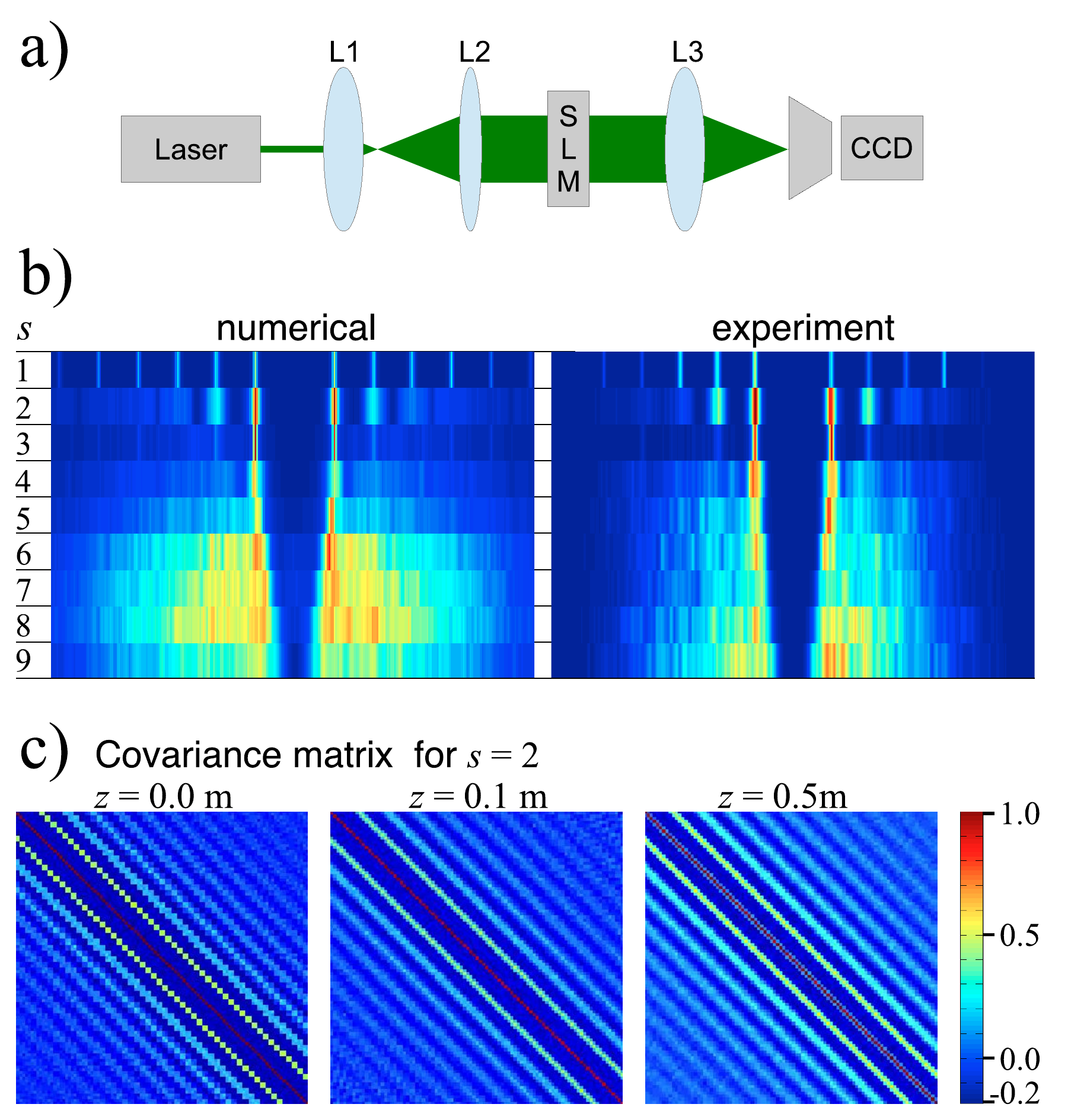}
	\caption[One dimensional]{Probing the far field of spatially non-Markovian light: a) Experimental setup; b) numerical (left side) and experimental (right side) results, averaged over multiple realizations, for 1D non-Markovian phase imprint with 9 equidistant phase levels. Rows depict the k-space for $s$ values from $9$ (top row) to $1$ (bottom row); c) central part of covariance matrix $J$ of the complex field for $s=2$ at different propagation distances $z$.}
	\label{fig:1D}
\end{figure}

To translate the NMP to a phase imprint, each element is assigned a discrete phase level by multiplying the respective number with $2 \pi / L$, and a SLM (Hamamatsu LCOS-SLM x10468) is programmed with this phase information. The properties of a of a plane wave ($\lambda =$~808~nm) reflected from the SLM surface is then probed in the experimental setup depicted schematically in Fig.~\ref{fig:1D}(a), in which the far field is recoded with a CCD camera.

Applying a 1D NMP at the SLM, we observe in the far field a dark central region (Fig.~\ref{fig:1D}(b)). Since the phase levels in the experiment are chosen such that the sum of the complex field is zero, a small dark region in the center is to be expected - similar e.g. to the center of an optical vortex. However, we note that the dip is significantly larger and depends in its width, edge shape and depth on the parameter $s$: For $s=1$, which generates strictly periodic patterns, the width reaches from the $-1^{st}$ to the $1^{st}$ diffraction order. The width narrows for increasing $s$ until about half this size for $s=9=L$, which corresponds to a series of random permutations without overlap or 'shared' elements. Between these two extremes we observe a transition where with increasing $s$ the higher diffraction orders (corresponding to higher spatial frequencies) 'wash out' and build a relative homogeneous noise floor, normally a characteristic of spatially incoherent light. This transition is not linear, as can be seen in Fig.~\ref{fig:1D}(b); in particular for values of $s$ that are divisors of $L$ (e.g. $s=3$) the lower diffraction orders and the central dip appear much more pronounced.

Using the same parameters as in the numerical simulation shown in Fig.~\ref{fig:1D}(b,~left), we calculated the covariance matrix of the complex field at the SLM surface and after a free-space propagation over a distance of 0.1~m and 0.5~m (illustrated exemplary for $s=2$ at Fig.~\ref{fig:1D}(c)). This covariance matrix, also referred to as the complex mutual intensity function, or degree of coherence function~\cite{horstmeyer_markov_2012}, reveals the interesting statistics of this light: anti-correlation in the immediate short-range and periodic peaks of high correlation at mid-range, which get weaker at increasing distance from the diagonal (the higher the ratio $s/L$, the faster the peak intensity decays). Interestingly enough, these main features persist even as the beam propagates and the overall degree of coherence increases in accordance with the Gaussian Schell-model~(causing the broadening of the diagonal for $z=0.1$~m and $0.5$~m)~\cite{mandel1995optical}.

\begin{figure}
	\centering
	\includegraphics[width=0.99\linewidth,clip=true]{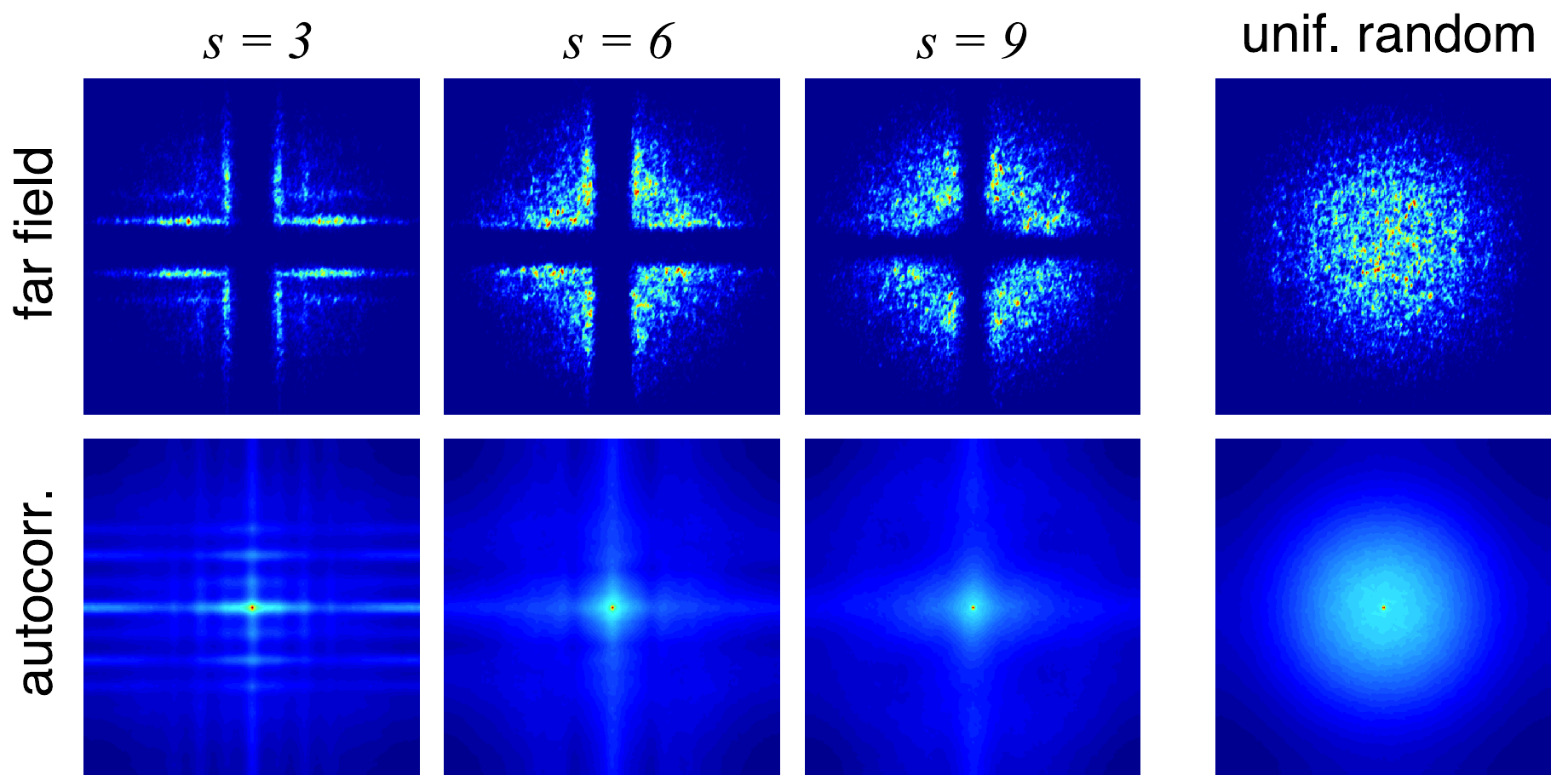}
	\caption[Two dimensional]{Sudokulight with 2D non-Markovian phase imprint. Top row (from left to right) depicts the far field measured with the setup illustrated in Fig.~\ref{fig:1D}(a) for NMPs with $s=3, 6$ and $9$, respectively, averaged over 100 realizations. Bottom row shows the far field intensity autocorrelation of the respective measurements. For comparison, the right column shows the respective results of a uniformly distributed random 9 level phase pattern imprinted on the beam.}
	\label{fig:2D}
\end{figure}

A similar behavior can be observed for the 2D case, where the pattern for non-Markovian phase imprint is generated by solving overlapping Sudoku puzzles. As depicted in Fig.~\ref{fig:2D}(top row), the far field features a dark cross, whose width and border shape again depend on the parameter $s$. The transition of structured to relative homogeneous autocorrelation traces of these measurements (Fig.~\ref{fig:2D}(bottom row)) illustrates the decreasing degree of spatial coherence for an increasing value of $s/L$. While the spatial coherence can thus be tuned by the choice of $s$, a comparison with the autocorrelation trace of a uniformly distributed random phase pattern (using the same 9 discreet phase levels as the Sudokulight; Fig.~\ref{fig:2D}(bottom row, right column)) makes clear that even the case $s=L$ is far from being spatially incoherent.

\begin{figure}
	\centering
	\includegraphics[width=0.9\linewidth,clip=true]{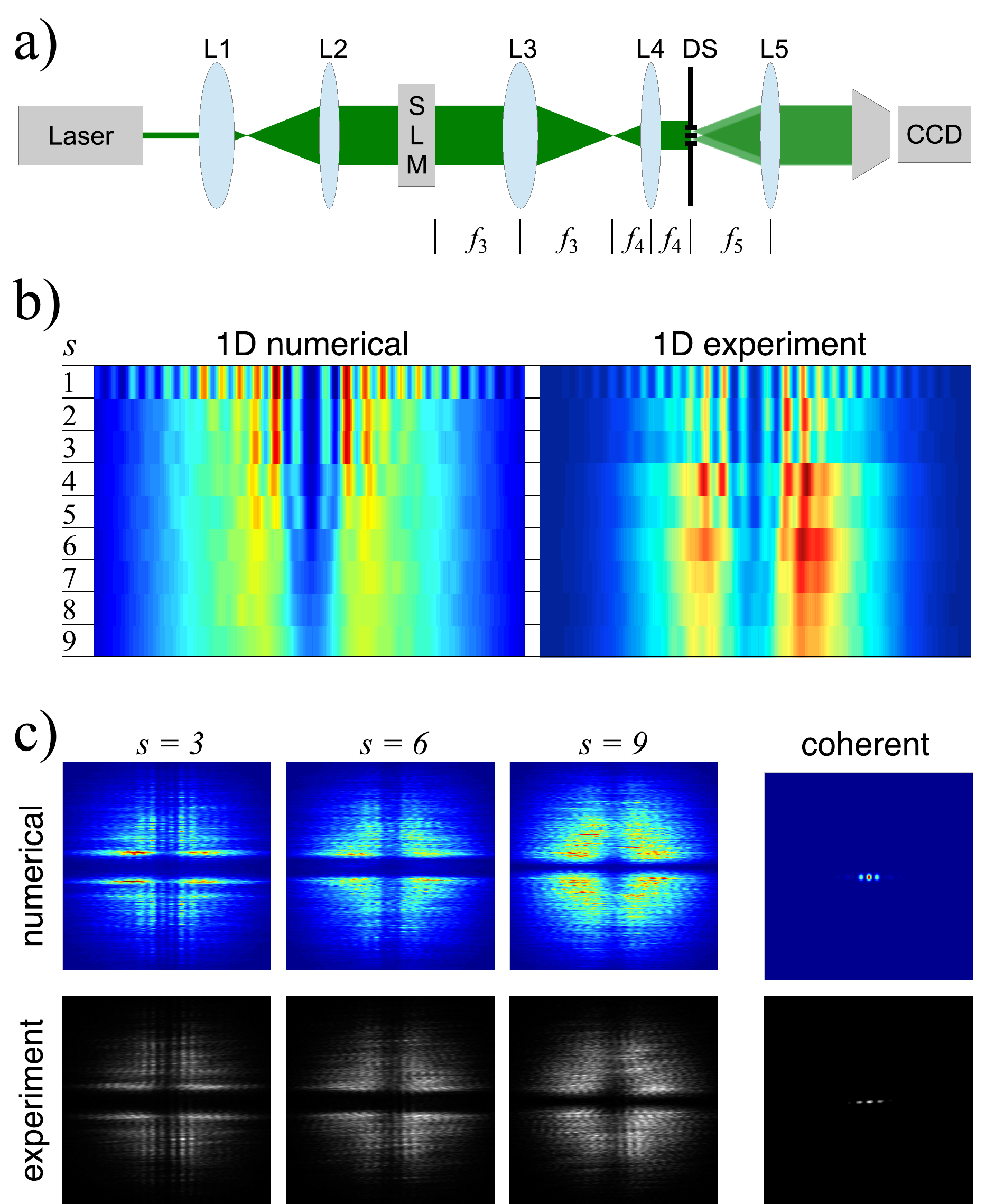}
	\caption[Double Slit]{Diffraction of Sudokulight on double slit. All results are averaged over 100 realizations with different NMPs. a) Experimental setup, employing a $4f$ imaging to reconstruct the phase imprint in the plane of the double slit with a magnification of $1/5$. b) Numerical (left side) and experimental (right side) diffraction patterns for 1D non-markovian light for increasin parameter $s$. Rows depicts the intensity of the diffraction pattern integrated along the slit orientation for $s=1$ (top) to $9$ (bottom). c) Top and bottom row show the numerical and experimental diffraction patterns obtained for Sudokulight with $s=3, 6$ and $9$, respectively (slits oriented vertically). The right column depicts the diffraction of a plane wave under the same conditions.}
	\label{fig:Young}
\end{figure}

These coherence characteristics also lead to an unusual diffraction pattern of Sudokulight in Young's classical double slit experiment~\cite{young_playingCardNcolourFringes_1804}. The fringe contrast at interference is commonly used as a measure for the coherence of light~\cite{goodman2000statistical}. To test the diffraction of Sudokulight on a double slit, we therefore employ the experimental setup schematically shown in Fig.~\ref{fig:Young}(a). A $4f$ imaging with a magnification of $1/5$ downscales the phase imprint at the plane of the double slit, such that the size of about 5 elements in the NMP correspond to the 100~$\mu$m slit width (slit distance 300~$\mu$m). The diffraction pattern is again recoded with a CCD camera.

 While incoherent light produces a Gaussian shaped intensity diffraction pattern perpendicular to the slit orientation, coherent light produces fringes as seen in Fig.~\ref{fig:Young}(c), right column. Partially coherent light is known to produce a linear combination of these two patterns, yet always with an intensity maximum at the center. Sudokulight, on the other hand, does feature an intensity minimum at the center, not only for the strictly periodic case $s=1$, but for all values of the tuning parameter $s$, both in the one-dimensional (Fig.~\ref{fig:Young}(b)) and the two-dimensional case (Fig.~\ref{fig:Young}(c), columns 1 to 3, all results are averaging over 100 realizations with different NMPs at fixed value of $s$). In the two-dimensional case, it is important to distinguish between the two pattern orientations. While the horizontal (perpendicular to the slit orientation) intensity modulation originates from the diffraction on the double slit, the vertical intensity modulation is simply the far field of the light along both slits. By comparing the two orientations in Fig.~\ref{fig:Young}(c), one notes the appearance of additional, vertical fringes due to interference of the diffracted light, which vanish for higher values of $s$. This decreasing fringe contrast for an increasing value of $s/L$, which can also be observed in the one dimensional case, confirms that the spatial coherence can be tuned with the parameter $s$.

We note that the pattern generation methods used in this work are but two possibilities among many to create spatially non-Markovian light via a SLM. To give just one more example, a NMP tiled with non-overlapping squares (width $w$), which contain a random permutation of numbers 1 to $w^2$, produces a donut shaped intensity pattern in the far field. In contrast to previous approaches generating non-Markovian light in time, the spatial analogue therefore offers a far greater freedom and flexibility (along with repeatability) in designing the stochastic and statistical properties of the light. Notwithstanding, the approach of using a SLM to imprint a non-Markovian phase can be easily extended to the time domain by rapidly reprogramming the SLM with a series of patterns which depend on their predecessors in a non-Markovian fashion. Different non-Markovian behaviors observed in natural systems can thus be simulated more accurately with the corresponding pattern generation method. Although any rule that correlates matrix elements beyond the immediate neighbor may produce a NMP, methods with a tuning parameter, such as the shift size $s$, allow for a deeper understanding of the impact of non-Markovian properties on the system of interest.

In conclusion, the generation of spatially non-Markovian light with tunable properties, as demonstrated in this work for one and two dimensions, lends itself to serve as testbed for the study of non-Markovian systems and dynamics. Light with a non-Markovian phase imprint shows very unusual behavior even in classical, linear experiments, such as the diffraction on a double slit. Further research has to show how this light interacts with nonlinear systems. Since it maintains it's particular spatial coherence properties when propagating in free-space, it might well serve as a building block to design specific, non-Markovian properties in the optical regime and then transfer them to other physical systems~\cite{Barreiro_pumpingTrappedIonsQuantumSim_2011,Paternostro_opticalCreatingNProbingEntanglement_2007}. Due to the flexibility in controlling the spatial coherence by the suggested pattern generation method and the easy, fast, and reproducible experimental implementation via a spatial light modulator, the approach presented in this work may become a useful tool in the search of a deeper understanding of non-Markovian processes.

This work was supported by grants from the CAPES-Weizmann cooperation program and CNPq (Universal grant 483983/2013-6).

\bibliography{sudokulight}{}
\bibliographystyle{plain}
\end{document}